\documentclass[aps,prd,twocolumn,superscriptaddress,nofootinbib]{revtex4-2}

\newcommand{\jcap}{J. Cosmol. Astropart. Phys. }
\newcommand{\apjs}{Astrophys. J. Suppl. Ser. }
\newcommand{\apjl}{Astrophys. J. Lett.}
\newcommand{\physrep}{Phys. Rep. }
\newcommand{\aap}{Astron. Astrophys. }

\newcommand{\mnras}{Mon. Not. R. Astron. Soc. }

\newcommand{\nphysa}{Nucl. Phys. A }
\newcommand{\apss}{Astrophys. Space Sci. }
\usepackage{color}
\usepackage{amsmath}

\usepackage{graphicx}
\usepackage{slashed}
\newcommand{\eda}{\mathcal{E}}
\newcommand{\fla}{\mathcal{F}}

\begin{document}

\title{Two-dimensional models of core-collapse supernova explosions assisted by heavy sterile neutrinos}

\author{Kanji Mori}
\email[]{kanji.mori@nao.ac.jp}
\altaffiliation{Research Fellow of Japan Society for the Promotion of Science}
\affiliation{National Astronomical Observatory of Japan, 2-21-1 Osawa, Mitaka, Tokyo 181-8588, Japan}
\author{Tomoya Takiwaki}
\affiliation{National Astronomical Observatory of Japan, 2-21-1 Osawa, Mitaka, Tokyo 181-8588, Japan}
\author{Kei Kotake}
\affiliation{Research Institute of Stellar Explosive Phenomena, and Department of Applied Physics, Fukuoka University, 8-19-1 Nanakuma, Jonan-ku, Fukuoka-shi, Fukuoka 814-0180, Japan}
\affiliation{Institute for Theoretical Physics, University of Wroc\l aw, 50-204 Wroc\l aw, Poland}
\author{Shunsaku Horiuchi}
\affiliation{Center for Neutrino Physics, Department of Physics, Virginia Tech, Blacksburg, VA 24061, USA}
\affiliation{Kavli IPMU (WPI), UTIAS, The University of Tokyo, Kashiwa, Chiba 277-8583, Japan}

\date{\today}

\begin{abstract}
Core-collapse supernovae can be a copious source of sterile neutrinos, hypothetical particles that mix with active neutrinos. 
We develop two-dimensional stellar core-collapse models that incorporate the mixing between tau neutrinos and heavy sterile neutrinos---those with the mass of 150--200\,MeV---to investigate  signatures of sterile neutrinos in supernova observables. We find that the decay channel of a sterile neutrino into a pion and a tau neutrino can enhance the explosion energy and the synthesized nickel mass.  Although the inclusion of sterile neutrinos considered in this study slightly reduce the neutrino and gravitational-wave signals, 
we find that they are still detectable for a Galactic event. Furthermore, we point out that if sterile neutrinos are as massive as $\sim200$\,MeV, they produce high-energy tau antineutrinos with energies of $\sim80$\,MeV, the detection of which can be a smoking signature of the sterile neutrinos and where Hyper-Kamiokande should play a pivotal role.

\end{abstract}

\maketitle

\section{Introduction}

In the Standard Model of particle physics, neutrinos are treated as massless left-handed fermions. However, observations of neutrino oscillations \cite[e.g.,][]{2000PhRvL..85.3999F,2001PhRvL..87g1301A} reveal that the neutrino masses are nonzero. Although the origin of the neutrino masses is under debate, introducing heavy right-handed particles called sterile neutrinos \cite{1968JETP...26..984P,2021PhR...928....1D} naturally leads to finite masses of active neutrinos through the  seasaw mechanism \cite{1977PhLB...67..421M,1980PhRvL..44..912M,1980PThPh..64.1103Y}. 

Apart from the theoretical motivation above, there are experimental hints of sterile neutrinos. For example, electron antineutrino fluxes from reactors are anomalously smaller than theoretical expectations \cite{2011PhRvC..84b4617H,2011PhRvC..83e4615M,2016ARNPS..66..219H}. This deficit could be explained by oscillations into sterile neutrinos \cite{2011PhRvD..83g3006M}. Also, sterile neutrino with mass $\sim$keV work as a candidate of dark matter \cite{1994PhRvL..72...17D,1999PhRvL..82.2832S,2019PrPNP.104....1B}. Interestingly, x-ray observations of galaxy clusters and galaxies with the XMM-Newton and Chandra satellites detected a 3.5\,keV line, which can be interpreted as a signature of the radiative decay of 7.1\,keV-mass sterile neutrino dark matter \cite{2014ApJ...789...13B,2014PhRvL.113y1301B,2015PhRvL.115p1301B}. Although the sterile neutrino interpretation for these signatures is under debate \cite[e.g.,][]{Horiuchi:2013noa,Abazajian:2014gza,2014PhRvL.112t2501H,2016MNRAS.456.4346H,2016ARNPS..66..219H,2019PrPNP.104....1B,2020Sci...367.1465D,Roach:2022lgo}, these studies motivate experimental and astronomical explorations of the particle. 

Core-collapse supernovae are a useful laboratory for sterile neutrinos because neutrinos play essential roles in the supernova explosion mechanism. Sterile neutrinos can be produced in supernova events through mixing with active neutrinos and can affect the energy transfer inside. Indeed, the effects of sterile neutrinos on supernovae have been investigated for various sterile neutrino masses $m_\mathrm{s}=\mathcal{O}(\mathrm{eV})$ \cite{1991NuPhB.358..435K,1992A&A...254..121P,1993APh.....1..165R,1997PhRvD..56.1704N,1999PhRvC..59.2873M,2000PhRvD..61l3005C,2003APh....18..433F,2006PhRvD..73i3007B,2007PhRvD..76l5026K,2012JCAP...01..013T,2014PhRvD..89f1303W}, $\mathcal{O}(\mathrm{keV})$ \cite{1994PhLB..323..360S,2001PhRvD..64b3501A,2006PhRvD..74l5015H,2007PhRvD..76h3516H,2011PhRvD..83i3014R,2014PhRvD..90j3007W,2016IJMPA..3150137W,2019PhRvD..99d3012A,2019JCAP...12..019S,2020JCAP...08..018S,2022PhRvD.106a5017S,2023PhRvD.108f3025R}, and $>\mathcal{O}(\mathrm{MeV})$ \cite{2000NuPhB.590..562D,2009PhLB..670..281F,2015PhLB..751..209A,2018PhRvD..98j3010R,2020JCAP...01..010M,2023arXiv230905860C,2023arXiv231100033C,2023arXiv231213627A}. Although these studies have gradually refined the microphysics, they nonetheless have only adopted spherically-symmetric supernova models. This is despite the fact that, multi-dimensionality is essential to the supernova explosion mechanism. Also, it is not possible to predict gravitational wave signals with one-dimensional models. In this study, we hence develop two-dimensional supernova models coupled with sterile neutrinos.

We focus on a heavy sterile neutrino model with $m_\mathrm{s}=150$-200\,MeV and Lagrangian \cite{2009PhLB..670..281F,2023arXiv230905860C},
\begin{eqnarray}
\mathcal{L}=\mathcal{L}_\mathrm{SM}+i\bar{\nu}_\mathrm{s}{\partial\!\!\! /}\nu_\mathrm{s}-y_\nu \bar{L}\tilde{H}\nu_\mathrm{s}+\frac{m_\mathrm{s}}{2}\bar{\nu}_\mathrm{s}^c\nu_\mathrm{s}+\mathrm{h.c.},
\end{eqnarray}
where $\mathcal{L}_\mathrm{SM}$ is the Standard Model Lagrangian, $y_\nu$ is a Yukawa coupling constant, $H$ is the SU(2)$_L$ Higgs doublet, and $L$ is the lepton doublet. The Yukawa coupling term leads to a Dirac mass term below the electroweak scale and diagonalization of the Dirac and Majorana mass terms provides the masses of active and sterile neutrinos through the seesaw mechanism. We assume that sterile neutrinos mix with $\nu_\tau$,
\begin{eqnarray}
\nu_\tau&=&\cos\theta_{\tau4}\nu_1+\sin\theta_{\tau4}\nu_2\nonumber\\
\nu_\mathrm{s}&=&-\sin\theta_{\tau4}\nu_1+\cos\theta_{\tau4}\nu_2,
\end{eqnarray}
where $\nu_1$ and $\nu_2$ are the mass eigenstates and $\theta_{\tau4}$ is the mixing angle. Heavy sterile neutrinos that mix with $\nu_\mathrm{e}$ and $\nu_\mu$ have been strongly excluded by terrestrial experiments \cite{2015PhRvD..91e2001A,2018PhLB..778..137C,2019PhRvD.100e2006A}, but mixing with $\nu_\tau$ is less constrained. Core-collapse supernovae can provide a unique way to constrain the mixing angle $\theta_{\tau4}$ between $\nu_\tau$ and $\nu_\mathrm{s}$. Such a constraint has been set from SN 1987A \cite{2009PhLB..670..281F,2020JCAP...01..010M}. If sterile neutrinos exist, they can freely escape from the proto-neutron star and lead to an additional energy loss. Since this would affect the neutrino signals observed from SN 1987A, the mixing angle should be  $\sin^2\theta_{\tau4}\lesssim2\times10^{-6}$ for $m_\mathrm{s}=200$\,MeV \cite{2009PhLB..670..281F,2020JCAP...01..010M}. In our simulations, we adopt sterile neutrino parameters that do not violate this constraint based on SN 1987A energy-loss argument.

This paper is organized as follows. In Section II, we explain our method to implement the transport of sterile neutrinos in supernova simulations. In Section III, we show the results of our simulations and discuss the effect of sterile neutrinos. In Section IV, we summarize and discuss the implication of our results. Throughout, we use natural units with $\hbar=c=k=G=1$, where $\hbar$ is the Planck constant, $c$ is the speed of light, $k$ is the Boltzmann constant, and $G$ is the gravitational constant.

 \section{Method}
 \label{sec:method}

In this work, we consider a sterile neutrino model proposed in Ref.~\cite{2009PhLB..670..281F}. In a supernova core, a heavy sterile neutrino $\nu_\mathrm{s}$ is produced by the pair annihilation of $\nu_\mu$ and $\nu_\tau$. The processes considered in our calculation is tabulated in Refs.~\cite{2009PhLB..670..281F,2020JCAP...01..010M}. Although the annihilation processes of $\nu_\mathrm{e}$-$\bar{\nu}_\mathrm{e}$ pairs and electron-positron pairs  can produce $\nu_\mathrm{s}$ as well, their contribution would be subdominant because these species are degenerate in the supernova core. We adopt the numerical result in Ref.~\cite{2009PhLB..670..281F} for the cooling rate induced by the sterile neutrino production,
\begin{eqnarray}
    &Q_\mathrm{cool}&=3\times10^{34}\,\mathrm{erg\,cm^{-3}\,s^{-1}}\times\nonumber\\
    &&\left(\frac{\sin^2\theta_{\tau4}}{5\times10^{-8}}\right)\left(\frac{T}{35\,\mathrm{MeV}}\right)^{7.2}\exp\left(-\frac{m_\mathrm{s}}{T}\right),\label{Qcool}
\end{eqnarray}
where $T$ is the temperature. 

In this study, we explore sterile neutrino mass $m_\mathrm{s}$ of 150 and 200\,MeV. This range of mass is particularly interesting because the sterile neutrino is heavier than the neutral pion, whose mass is $m_\pi\approx135$\,MeV \cite{2022PTEP.2022h3C01W}. In this case, the decay channel $\nu_\mathrm{s}\rightarrow\nu_\tau\pi^0\rightarrow\nu_\tau\gamma\gamma$ and its charge conjugate reaction are kinematically allowed, and its lifetime can be estimated as \cite{2000NuPhB.590..562D,2009PhLB..670..281F,2018PhRvD..98j3010R},
\begin{eqnarray}
    \tau_\mathrm{s}&=&\frac{16\pi}{G_\mathrm{F}^2m_\mathrm{s}(m_\mathrm{s}^2-m_\pi^2)f_\pi^2\sin^2\theta_{\tau4}}\\
    &\approx&66\,\mathrm{ms}\left(\frac{\sin^2\theta_{\tau4}}{5\times10^{-8}}\right)^{-1}\left(\frac{m_\mathrm{s}}{200\,\mathrm{MeV}}\right)^{-3}\left(\frac{0.54}{1-\frac{m_\pi^2}{m_\mathrm{s}^2}}\right),\nonumber
\end{eqnarray}
where $G_\mathrm{F}$ is the Fermi coupling constant and $f_\pi\approx131$\,MeV. If $\nu_\mathrm{s}$ is heavier, other decay channels such as $\nu_\mathrm{s}\rightarrow\nu_\tau\pi^+\pi^-$ would appear. These processes will shorten the lifetime of sterile neutrinos, but investigation of their effects is beyond the scope of this study.

In order to simulate the transport of sterile neutrinos, we solve the zeroth angular moment of the Boltzmann equation \cite{2018PhRvD..98j3010R},
\begin{eqnarray}
\frac{\partial \eda }{\partial t}
+\nabla\cdot \mathbf{\fla}
=Q_\mathrm{cool}-\kappa\eda,\label{Boltz}
\end{eqnarray}
where $\eda$ is the $\nu_\mathrm{s}$ energy per unit volume, $\mathbf{\fla}$ is the $\nu_\mathrm{s}$ energy flux, and $\kappa=1/\tau_\mathrm{s}c$ is the opacity for the $\nu_\mathrm{s}$ decay. We adopt the ray-by-ray approximation, in which the non-radial propagation of sterile neutrinos is not considered. We adopt $\fla=c \eda$ as the closure relation because sterile neutrinos only feebly interact with the surrounding material. The choice of the closure relation could be justified by the feeble interaction of sterile neutrinos. On the other hand, Ref.~\cite{2018PhRvD..98j3010R} adopts the two-moment method to solve the Boltzmann equation for sterile and active neutrinos.  The nonlinear nature of the problem prevents one from predicting the consequences of different approximations.  It is hence desirable to perform comparative studies on sterile neutrino transport methods, which are beyond the scope of this study.

When the density of the surrounding material is high enough, the active neutrinos produced by the sterile neutrino decay are absorbed locally and contribute to heating. On the other hand, when the density is low, the active neutrinos will freely escape. In this study, we follow the prescription in Ref.~\cite{2018PhRvD..98j3010R} to calculate the efficiency of the additional heating induced by sterile neutrino decay. If the density is higher than $10^{12}$\,g\,cm$^{-3}$, the active neutrinos are assumed to be trapped and the heating rate is $Q_\mathrm{heat}=\kappa\eda$. When density is lower than $10^{10}$\,g\,cm$^{-3}$, the heating rate is calculated as $Q_\mathrm{heat}=\kappa\eda x_{\gamma\gamma}$, where $x_{\gamma\gamma}=0.5(1+m_\pi^2/m_\mathrm{s}^2)$. When the density is intermediate, $Q_\mathrm{heat}$ is linearly interpolated between the two regimes. Once $Q_\mathrm{cool}$ and $Q_\mathrm{heat}$ are obtained, sterile neutrinos are  coupled with the hydrodynamics through 
\begin{eqnarray}
\left( \frac{\partial e_{\mathrm{int}}}{\partial t}\right)_{\nu_s}= Q_{\mathrm{heat}} -Q_{\mathrm{cool}},
\end{eqnarray}
where $\left(\frac{\partial e_{\mathrm{int}}}{\partial t}\right)_{\nu_s}$ is the contribution of $\nu_s$ in the source term of the energy equation.

We implement the transfer of sterile neutrinos described above in the \texttt{3DnSNe} code \cite{2016MNRAS.461L.112T} to simulate the stellar core collapse with axisymmetric geometry. The transfer of active neutrinos is treated with a three-flavor isotropic diffusion source approximation \cite{2009ApJ...698.1174L,2014ApJ...786...83T,2018ApJ...853..170K}. We also solve the $\alpha$ network that includes 13 nuclei to consider the nuclear energy generation and nucleosynthesis \cite{2014ApJ...782...91N}.  We adopt the nuclear equation of state in Ref.~\cite{1991NuPhA.535..331L} with $K=220$\,MeV. The spatial resolution of our two-dimensional simulations is $n_r\times n_\theta=512\times 128$. The simulated region is a sphere with radius of $5000$\,km. The progenitor model is the non-rotating $20M_\odot$ solar metallicity star in Ref.~\cite{2007PhR...442..269W}. 

\begin{table*}[]
\begin{tabular}{ccc|cccccc}
Model&$m_\mathrm{s}$ [MeV] & $\sin^2\theta_{\tau4}$& MFP [$10^4$ km] & $t_\mathrm{pb,\;2000}$ [ms]& $E_\mathrm{diag}$ [$10^{51}$\,erg] &$M_\mathrm{Ni}/M_\odot$&$M_\mathrm{PNS}/M_\odot$&$E_\mathrm{s}$ [$10^{51}$\,erg]\\\hline\hline
   Standard&$-$   & 0  & $-$ &  395   & 0.36 & 0.077 & 1.83 & 0                              \\\hline

$(150,\;6)$&150   & $6\times10^{-8}$        & 11.5 &   370         &   0.61&0.086                                   &  1.82 &   8.77           \\
$(150,\;10)$&150   & $1\times10^{-7}$        & 6.92 &   360                                 &  0.76&  0.084            & 1.82  & 12.2   \\
$(150,\;20)$&150   & $2\times10^{-7}$         & 3.46 &    321                                 &  2.02&  0.089    & 1.80  & 15.3          \\
\hline
$(200,\;6)$&200   & $6\times10^{-8}$        & 1.64 &   362                                 &  0.59&  0.088   &1.82&     1.64          \\
$(200,\;10)$&200   & $1\times10^{-7}$        & 0.99 &    345                                 &  0.87&  0.085        &1.81&1.85         \\
$(200,\;20)$&200   & $2\times10^{-7}$        & 0.49 &   316                                 &  2.26&  0.105             &1.78&1.66      \\

\end{tabular}
\caption{Parameters for the supernova models developed in this study and results. We adopt a ``Standard" model without sterile neutrinos and six models with sterile neutrinos. Each of the latter models is designated by a pair of two integers that indicates the mass and mixing angle, i.e., $(m_\mathrm{s}$ [MeV],\;$\sin^2\theta_{\tau4}/10^{-8})$. The fourth column shows the mean free path (MFP) of sterile neutrinos. The fifth  column shows the post bounce time $t_\mathrm{pb,\;2000}$ at which the bounce shock reaches the radius $2000$\,km when angle-averaged. The sixth column shows the diagnostic explosion energy defined in Eq.~(\ref{Ediag}). The seventh column shows the synthesized nickel mass defined in Eq.~(\ref{Nimass}). The eighth column shows the additional heating outside the simulated region based on Eq.~(\ref{Es_eq}). $E_\mathrm{diag}$, $M_\mathrm{Ni}$, $M_\mathrm{PNS}$, and $E_\mathrm{s}$ are all evaluated at $t_\mathrm{pb}=t_\mathrm{pb,\;2000}$.}
\label{table}
\end{table*}
\section{Results}
In this work, we develop six models with sterile neutrinos and one model without sterile neutrinos (the ``Standard'' model). The $\nu_\mathrm{s}$ mass is assumed to be 150 or 200\,MeV and the mixing angle is $\sin^2\theta_{\tau4}=6\times10^{-8},\;1\times10^{-7}$, or $2\times10^{-7}$. The characteristics of the models are tabulated in Table \ref{table}. 

Note that while our chosen sterile neutrino parameter evade the constraint in Ref.~\cite{2020JCAP...01..010M} that is based on the supernova energy loss argument, they are in the excluded region based on the explosion energy of low-energy supernovae \cite{2023arXiv230905860C,2023arXiv231100033C}. However, the explosion energy constraint is based on a post-process calculation of the sterile neutrinos' effects, and thus it is still worthwhile to investigate the dynamical effects of heavy sterile neutrinos in our numerical implementation. 
 
In this section, we discuss the properties of our models. Each model is designated by a pair of two numbers, $(m_\mathrm{s}/1\,\mathrm{MeV},\;\sin^2\theta_{\tau4}/10^{-8})$, throughout the paper.

 \begin{figure}
  \centering
    \includegraphics[width=9cm]{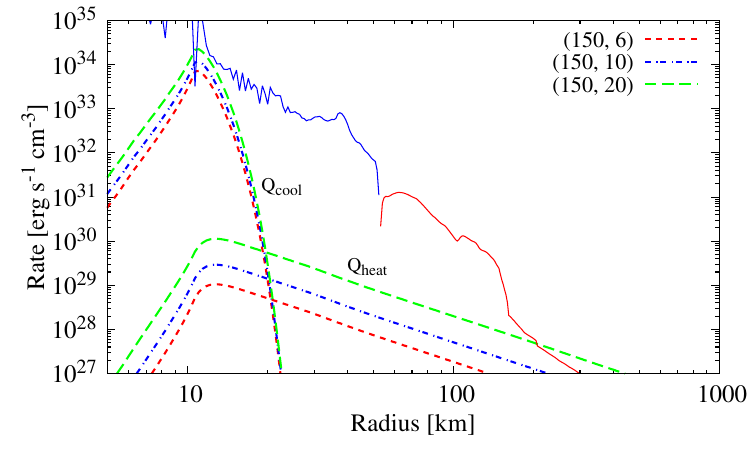}
  \includegraphics[width=9cm]{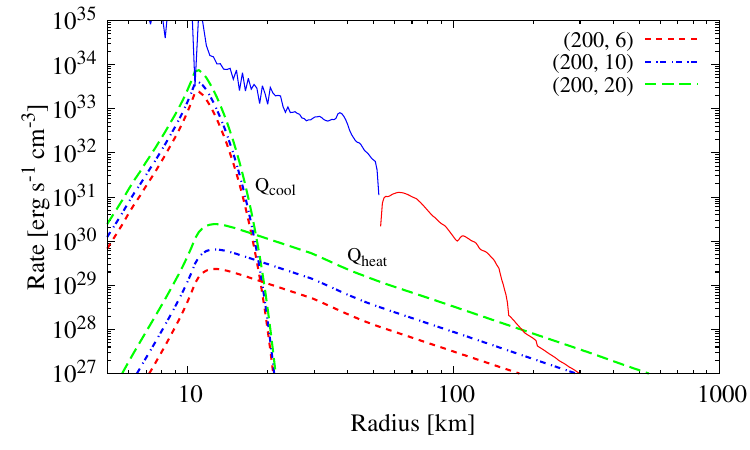}
  \caption{The angular-averaged radial profile of the cooling rate $Q_\mathrm{cool}$ and the heating rate $Q_\mathrm{heat}$ induced by sterile neutrinos, with $m_\mathrm{s}=150$\,MeV (upper) and 200\,MeV (lower) but different mixing angles. The solid lines indicate the net heating/cooling rate by active neutrinos, $|Q_\mathrm{active}|$, as defined in Eq.~(\ref{Qactive}). Here the red and blue parts represent $Q_\mathrm{active}>0$ and $Q_\mathrm{active}<0$, respectively. All shown for $t_\mathrm{pb}=0.2$\,s.}
  \label{Q}
 \end{figure}

 \begin{figure}
  \centering
  \includegraphics[width=8cm]{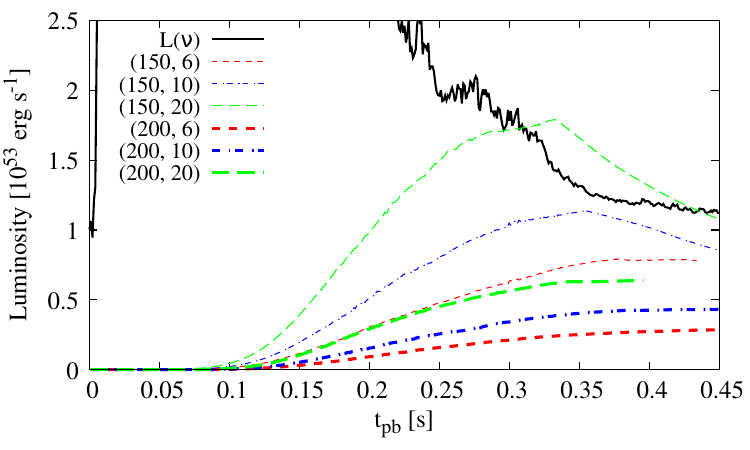}
  \caption{The active neutrino luminosity (thick solid black) and the sterile neutrino luminosity $L_\mathrm{s}|_{r=500\,\mathrm{km}}$ (as labeled), all  evaluated at $r=500$\,km as functions of $t_\mathrm{pb}$. 
  The thin lines are for the models with $m_\mathrm{s}=150$\,MeV while the thick lines are for the models with $m_\mathrm{s}=200$\,MeV. Different line styles represent different mixing angles.}
  \label{Ls}
 \end{figure}

\subsection{Additional Cooling and Heating}

Sterile neutrino contributes to both cooling and heating in addition to the active neutrinos.
Figure~\ref{Q} shows the radial profile of $Q_\mathrm{cool}$  at the postbounce time $t_\mathrm{pb}=0.2$\,s. The cooling rate $Q_\mathrm{cool}$ reaches a peak at $r\approx10$\,km. This is because the temperature reaches the highest value, $\sim45$\,MeV, at this radius and $Q_\mathrm{cool}$ is very sensitive to the temperature as shown in Eq.~(\ref{Qcool}). Figure~\ref{Q} also shows the radial profile of $Q_\mathrm{heat}$, which is obtained by solving Eq.~(\ref{Boltz}). When the sterile neutrino mass and the radius are fixed, $Q_\mathrm{heat}$ is approximately proportional to $\sin^4\theta_{\tau4}$ because both  $Q_\mathrm{heat}$ and $\kappa$ are proportional to $\sin^2\theta_{\tau4}$. 

Figure~\ref{Q} also shows the net heating and cooling rate
\begin{eqnarray}
Q_\mathrm{active}=Q_\mathrm{heat,\,active}-Q_\mathrm{cool,\,active}\label{Qactive}
\end{eqnarray}
induced by active neutrinos. Here the ``Standard" model was used to estimate the $Q_\mathrm{active}$ profile. We can see that the gain radius, where cooling balances with heating, is located at $r\approx53$\,km at $t_\mathrm{pb}=0.2$\,s. At this radius, the contribution of sterile neutrinos on heating and cooling is smaller than the contribution of active neutrinos. Hence, the gain radius is not affected by sterile neutrinos. In addition, the figure shows that $Q_\mathrm{cool}$ for $m_\mathrm{s}=200$\,MeV is smaller than that for $m_\mathrm{s}=150$\,MeV when the mixing angle is fixed. This is because of the Boltzmann factor in Eq.\,(\ref{Qcool}). The difference in $Q_\mathrm{heat}$ is, however, less significant. This can be attributed to the longer lifetime for lighter sterile neutrinos, which leads to a smaller opacity. We note that $Q_\mathrm{heat}$  in the models with $\sin^2\theta_{\tau4}=2\times10^{-7}$ reaches $\sim1\%$ of $Q_\mathrm{heat,\,active}$. The values are so high that they could significantly affect supernova dynamics.

In Fig.~\ref{Ls}, we show the luminosity $L_\mathrm{s}|_{r=500\,\mathrm{km}}$ of sterile neutrinos evaluated at $r=500$\,km. When $m_\mathrm{s}$ is fixed, larger mixing angles lead to higher luminosities. When the mixing angle is fixed, the luminosities for the models with $m_\mathrm{s}=150$\,MeV are $\sim3$--4 times higher than those for the models with $m_\mathrm{s}=200$\,MeV, because the production of heavy sterile neutrinos is suppressed by the Boltzmann factor.  

Figure~\ref{Ls} also shows the total active neutrino luminosity for the Standard model without sterile neutrinos, i.e., $L_\nu=L_{\nu_e}+L_{\bar{\nu}_e}+4L_{\nu_X}$, which is also evaluated at $r=500$\,km. We can see that $L_\mathrm{s}$ is lower than $L_\nu$ in almost all models, with the exception of the $(150,\,20)$ model, which is our highest mixing angle model and also the mass is low enough so as to not be strongly Boltzmann suppressed. A secondary effect in the $(150,\,20)$ model is that the proto-neutron star cooling induced by the sterile neutrino production is so efficient that the temperature at the region where sterile neutrinos are produced is significantly lower. As a result, the sterile neutrino luminosity starts decreasing quickly from post-bounce time $t_\mathrm{pb}\sim0.34$\,s.

 \begin{figure}
  \centering
  \includegraphics[width=8cm]{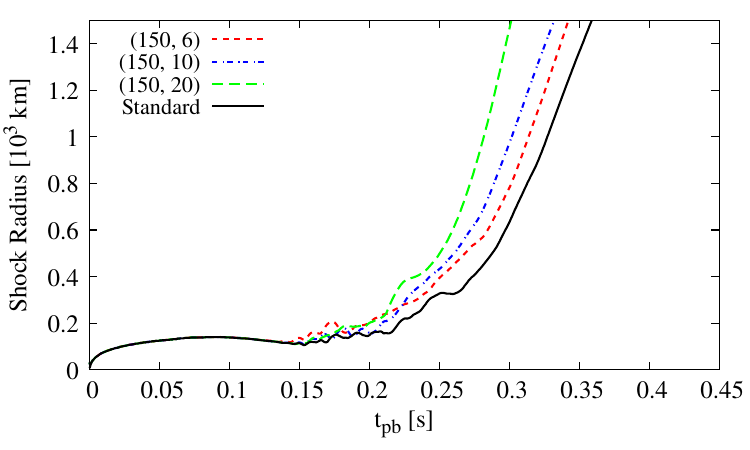}
   \includegraphics[width=8cm]{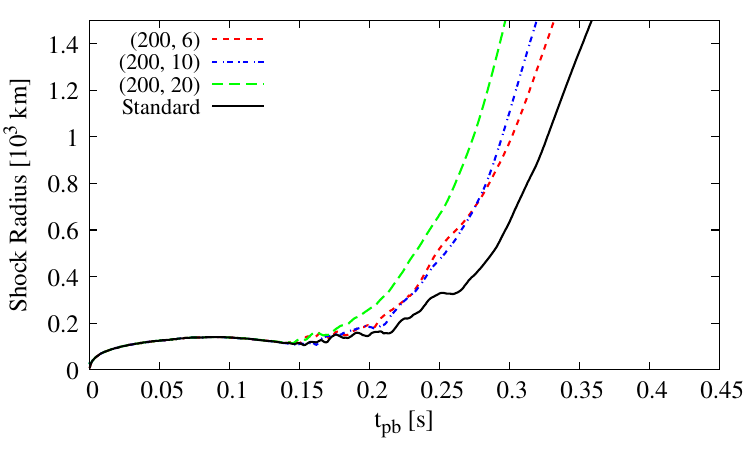}
  \caption{The angle-averaged radius of the bounce shock as functions of post-bounce time $t_\mathrm{pb}$. The upper panel is for the models with $m_\mathrm{s}=150$\,MeV and the lower panel is for $m_\mathrm{s}=200$\,MeV. The solid line shows the result for the Standard model for comparison.}
  \label{rsh}
 \end{figure} 

 \begin{figure}
  \centering
  \includegraphics[width=8cm]{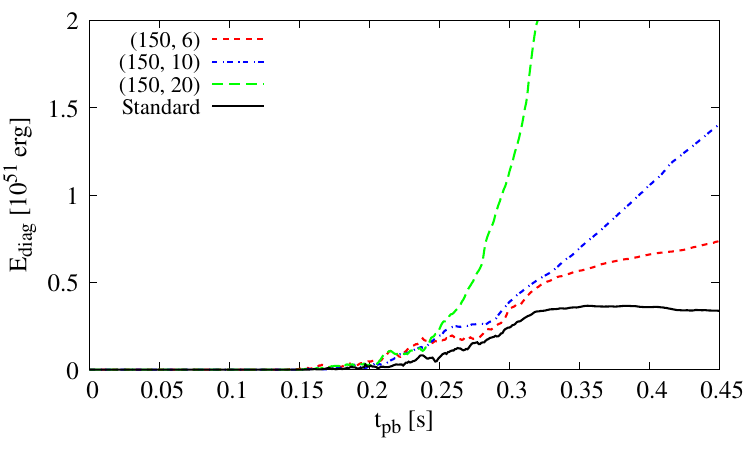}
   \includegraphics[width=8cm]{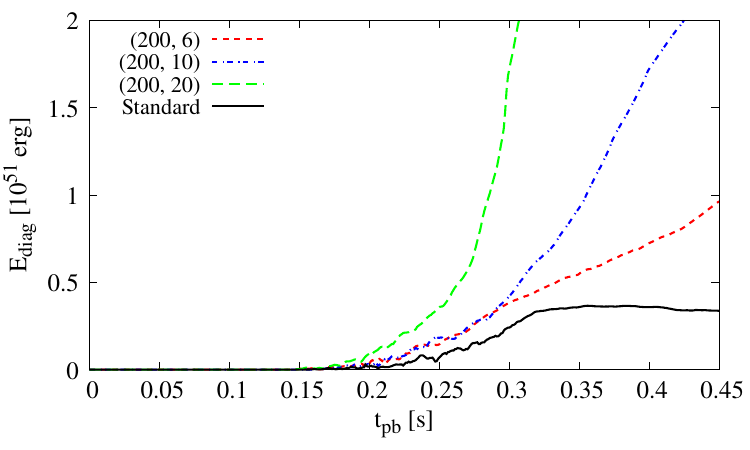}
  \caption{The diagnostic energy of explosion $E_\mathrm{diag}$ as functions of post-bounce time $t_\mathrm{pb}$.  The upper panel is for the models with $m_\mathrm{s}=150$\,MeV and the lower panel is for $m_\mathrm{s}=200$\,MeV. The solid line shows the result for the Standard model for comparison.}
  \label{Eexp}
 \end{figure}

  \begin{figure}
  \centering
  \includegraphics[width=8cm]{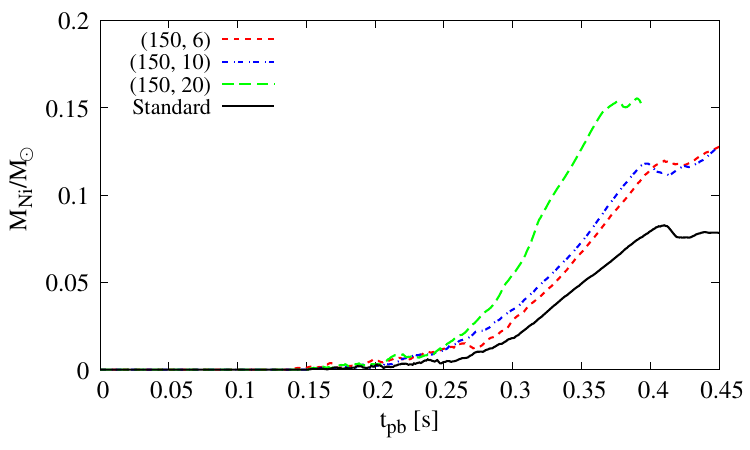}
   \includegraphics[width=8cm]{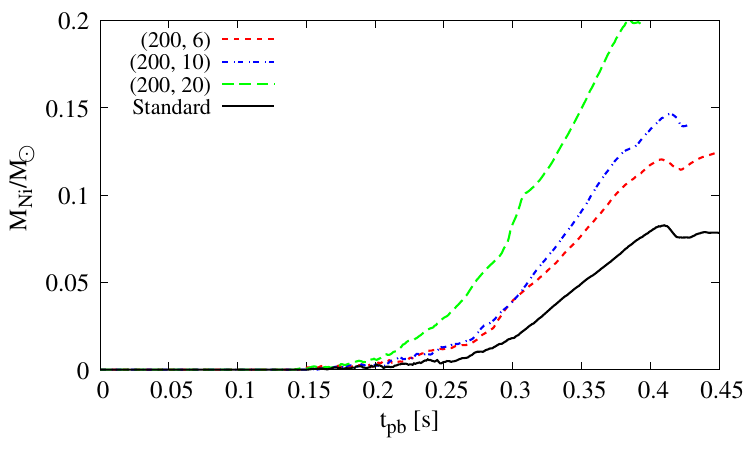}
  \caption{The synthesized nickel mass $M_\mathrm{Ni}$ as a function of $t_\mathrm{pb}$.  The upper panel is for the models with $m_\mathrm{s}=150$\,MeV and the lower panel is for the ones with $m_\mathrm{s}=200$\,MeV. The solid line shows the result for the Standard model  and the others are for the models with sterile neutrinos.}
  \label{Ni}
 \end{figure}

\subsection{Explosion Properties}

The additional cooling and heating rates shown in Fig.~\ref{Q} may significantly affect the explosion dynamics. Figure~\ref{rsh} shows the time evolution of the radius of the bounce shock. Since the models are two-dimensional, even the Standard model without sterile neutrinos achieves shock revival and a successful explosion is triggered. However, Fig.~\ref{rsh} shows that the additional heating induced by sterile neutrinos leads to faster expansion of the bounce shock. 

The effect of the additional heating can be seen in the explosion energy as well. Figure~\ref{Eexp} shows the diagnostic energy of explosion defined as,
\begin{equation}
      E_\mathrm{diag}=\int_D dV\left(\frac{1}{2}\rho v^2+e-\rho\Phi\right),\label{Ediag}
\end{equation}
where $\rho$ is the density, $v$ is the velocity, $e$ is the internal energy density, and $\Phi$ is the gravitational potential. The volume integral is performed in the region $D$ where the integrand is positive and the radial velocity is outward. Whereas $E_\mathrm{diag}$ stalls at $\sim0.4\times10^{51}$\,erg in the Standard model without sterile neutrinos, it continues to increase 
in the other models. Since $Q_\mathrm{heat}$ increases as a function of $\sin^2\theta_{\tau4}$, the explosion becomes more energetic when $\sin^2\theta_{\tau4}$ is larger. This is qualitatively consistent with the one-dimensional models of Ref.~\cite{2018PhRvD..98j3010R}. In particular, $E_\mathrm{diag}$ exceeds $1.5\times10^{51}$\,erg in the models with $\sin^2\theta_{\tau4}=1\times10^{-7}$ and $2\times10^{-7}$. In these cases, the predicted explosion energy is higher than the most frequent value, $0.6\times10^{51}$\,erg, adopted for observed Type II supernovae \cite{2022A&A...660A..41M}. This implies that the explosion energy is a useful observable to constrain the nature of sterile neutrinos \cite{2019PhRvD..99l1305S,2023arXiv230905860C,2023arXiv231100033C}.

It should be noted that the diagnostic energy of the explosion defined in Eq.~(\ref{Ediag}) cannot be directly compared with observed explosion energies, because $E_\mathrm{diag}$ does not consider effects of the overburden and sterile neutrinos that decay outside the simulated region. The overburden reduces the explosion energy by $\sim(0.2$--$0.3)\times10^{51}$\,erg from the diagnostic energy when the shock radius is in the range of 2000--5000\,km \cite{2023PhRvD.108f3027M}. On the other hand, even if the decay happens outside the shock radius, it enhances the internal energy in the envelope and mitigates the overburden effect, leading to an enhancement of the asymptotic value for the explosion energy.  The additional heating induced by sterile neutrinos which decay outside the simulated region enhances the explosion energy and can be estimated as,
\begin{eqnarray}
    E_\mathrm{s}=x_{\gamma\gamma}\int^t_0dt_\mathrm{pb}\left.L_{\rm s}\right|_{r=5000\,{\rm km}},\label{Es_eq}
\end{eqnarray}
where $L_\mathrm{s}|_{r=5000\,{\rm km}}$ is the sterile neutrino luminosity at radius $r=5000$\,km. As shown in Table I, $E_\mathrm{s}$ is between (2--20)$\times10^{51}$\,erg, which is significantly larger than $E_\mathrm{diag}$, in our parameter range. This implies that the explosion energy would grow beyond $E_\mathrm{diag}$ as the bounce shock propagates through the stellar envelope.

In Fig.~\ref{Ni} we show the $^{56}$Ni mass, 
\begin{eqnarray}
    M_\mathrm{Ni}=\int_D dVX_\mathrm{Ni}\rho,\label{Nimass}
\end{eqnarray}
where $X_\mathrm{Ni}$ is the $^{56}$Ni mass fraction and the volume integral is performed in the same region as the integral in Eq.~(\ref{Ediag}). The nickel mass saturates at $\sim0.07M_\odot$ in the standard model and becomes larger when the sterile neutrinos are considered. Nickel is mainly produced in the hot region where the temperature exceeds $\sim5\times10^9$\,K and complete silicon burning occurs \cite[e.g.,][]{2012ApJ...757...69U,2015ApJ...801...90P,2019MNRAS.483.3607S,2021ApJ...908....6S}. Due to the additional heating induced by sterile neutrinos, the volume of the hot region becomes larger and the resultant nickel masses can exceed $0.1M_\odot$. 

One can estimate the nickel mass in supernova events from their light curves \cite{1982ApJ...253..785A,1988ApJ...330..218W}. For example, the ejected nickel mass for SN 1987A is estimated to be $0.07M_\odot$ \cite{1988ApJ...330..218W}. Also, in a recent meta-analysis \cite{2019A&A...628A...7A} on observational papers of Type II supernovae, the median of the nickel mass is estimated to be $0.032M_\odot$  and the number of events with $M_\mathrm{Ni}>0.1M_\odot$ is less than 10\% of the total events. The comparison between these observations and our models implies that the models with $M_\mathrm{Ni}>0.1M_\odot$ overproduce nickel, although it would be necessary to perform more systematic simulations for a wide range of stellar parameters to obtain a solid constraint on the sterile neutrino. Furthermore, we discuss the limitations of two-dimensional simulations in Section \ref{ref:discuss}.

\subsection{Active Neutrinos}
\begin{figure*}
  \centering
  \includegraphics[width=5.5cm]{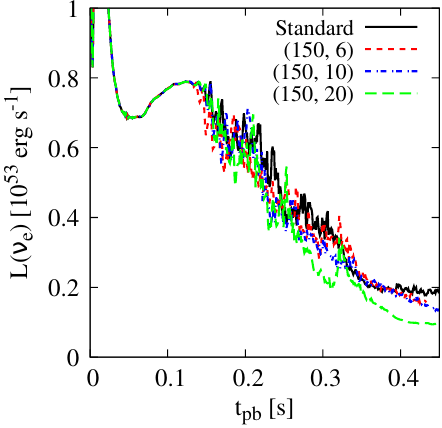}
  \includegraphics[width=5.5cm]{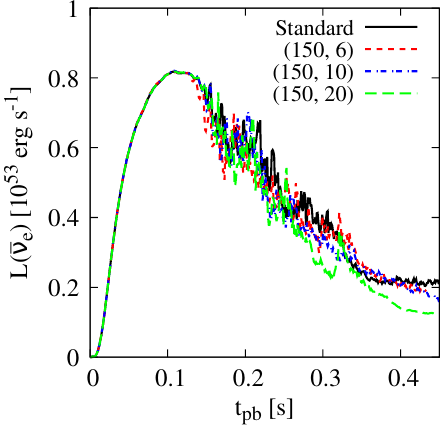}  \includegraphics[width=5.5cm]{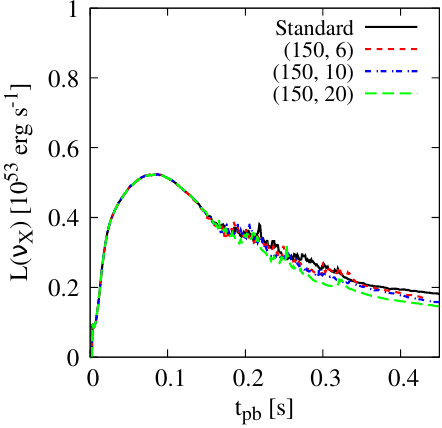}  \includegraphics[width=5.5cm]{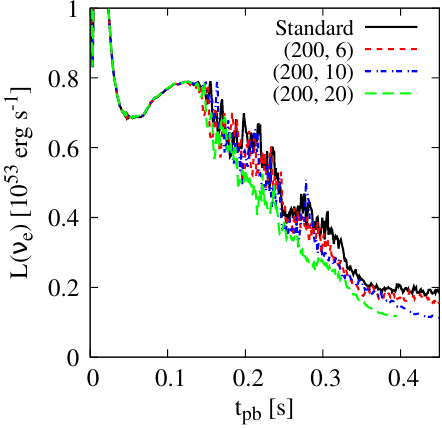}
  \includegraphics[width=5.5cm]{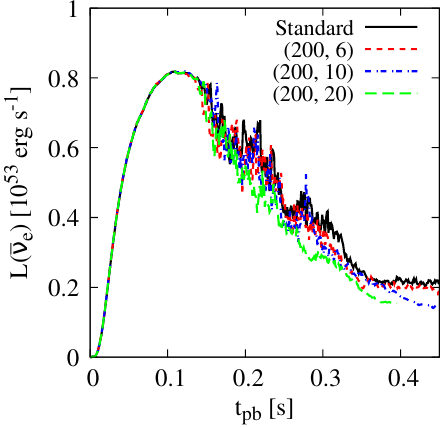}
  \includegraphics[width=5.5cm]{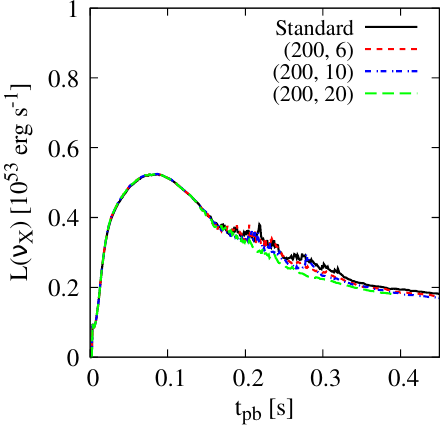}
  \caption{The luminosities of $\nu_e$, $\bar{\nu}_e$, and $\nu_X$ as functions of $t_\mathrm{pb}$. The contribution of decay (anti)neutrinos is not included.  The upper panels are for the models with $m_\mathrm{s}=150$\,MeV and the lower panels are for the ones with $m_\mathrm{s}=200$\,MeV. The solid line shows the result for the ``Standard" model  and the others are for the models with sterile neutrinos.}
  \label{Ln}
 \end{figure*}

\begin{figure*}
  \centering
  \includegraphics[width=5.5cm]{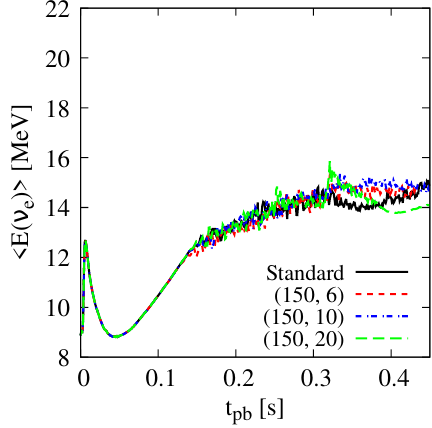}
  \includegraphics[width=5.5cm]{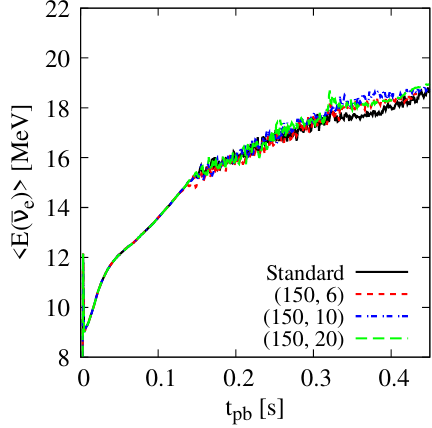}  \includegraphics[width=5.5cm]{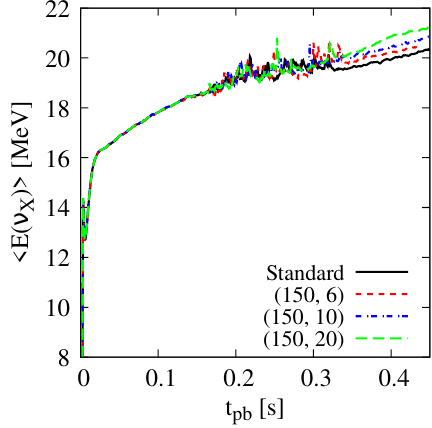}  \includegraphics[width=5.5cm]{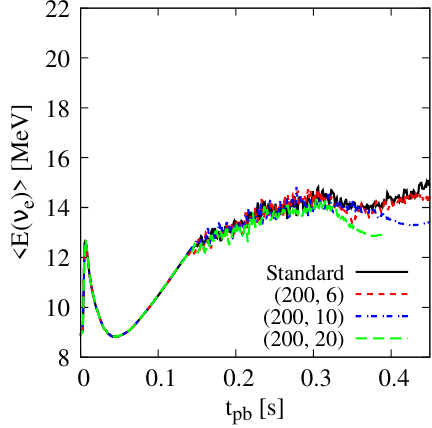}
  \includegraphics[width=5.5cm]{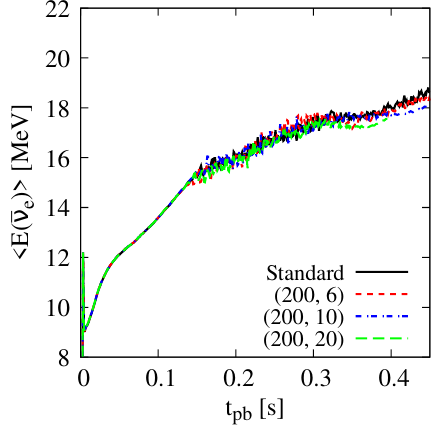}
  \includegraphics[width=5.5cm]{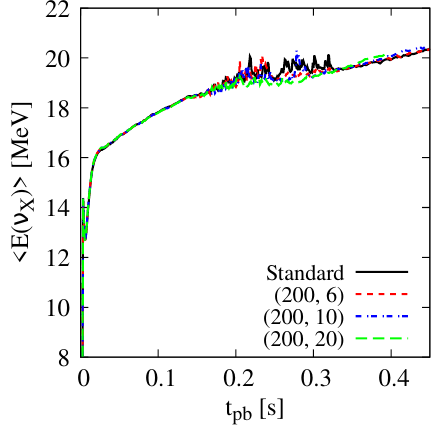}
  \caption{The mean energies of $\nu_e$, $\bar{\nu}_e$, and $\nu_X$ as a function of $t_\mathrm{pb}$.  The upper panels are for the models with $m_\mathrm{s}=150$\,MeV and the lower panels are for the ones with $m_\mathrm{s}=200$\,MeV. The solid line shows the result for the ``Standard" model  and the others are for the models with sterile neutrinos.}
  \label{En}
 \end{figure*}

The burst of active neutrino accompanying the core collapse can be detected by terrestrial detectors. The neutrino signal would provide information on the properties and dynamics in the collapsing core \cite[e.g.][]{2021MNRAS.506.1462N,2022ApJ...934...15S}. In particular, the energy transport induced by sterile neutrinos can affect the signals of active neutrinos.

Figure \ref{Ln} shows the luminosity of active neutrinos, namely electron neutrinos, electron (anti)neutrinos, and heavy lepton neutrinos. Note that the active neutrinos produced in sterile neutrino decays are not included. One can see that the luminosities of all flavors decrease as a function of the mixing angle. This is because the mass accretion is suppressed by the additional heating induced by the sterile neutrino decay. Similar decrease of active neutrino luminosities has been reported in two-dimensional supernova models with axionlike particles \cite{2023PhRvD.108f3027M} and QCD axions \cite{PhysRevD.106.063019} as well.

Figure \ref{En} shows the mean energy $\langle E\rangle$ of active neutrinos produced through the standard processes. Again, the active neutrinos produced in sterile neutrino decays are not included. Although the well-known hierarchy $\langle E(\nu_e)\rangle<\langle E(\bar{\nu}_e)\rangle<\langle E(\nu_X)\rangle$ is unchanged by sterile neutrinos, the absolute energies are affected by sterile neutrinos. The effect of sterile neutrinos is largely dependent on the sterile neutrino mass. When $m_\mathrm{s}=200$\,MeV, the mean energies are lower than those in the Standard model. This is because the temperature in the proto-neutron star is decreased by the additional cooling induced by the sterile neutrino production. This behavior is similar to the supernova models with axionlike particles \cite{2023PhRvD.108f3027M}. On the other hand, when $m_\mathrm{s}=150$\,MeV, the neutrino mean energies tend to become higher because of sterile neutrinos. This can be attributed to contraction of the proto-neutron star induced by the additional cooling effect. When the temperature is fixed, the sterile neutrino production rate is higher when sterile neutrinos are lighter because of the Boltzmann factor, as we can see in Eq.~(\ref{Qcool}). As a result, the cooling effect for the models with $m_\mathrm{s}=150$\,MeV becomes more significant compared with the models with $m_\mathrm{s}=200$\,MeV. The high cooling rate induces the proto-neutron star contraction and increases the central temperature $T_\mathrm{c}$. For example, the $(150,\;20)$ model shows $T_\mathrm{c}\approx30.8$~MeV at the post-bounce time $t_\mathrm{pb}=370$~ms, whereas the $(200,\;20)$ model shows $T_\mathrm{c}\approx25.5$~MeV. The increase in the mean energy is the most significant for heavy lepton neutrinos because $\nu_X$ interacts with the surrounding matter only through the neutral-current reactions and the radius of neutrinosphere is the smallest. This effect of the proto-neutron contraction is similar to that reported in supernova models with QCD axions \cite{PhysRevD.106.063019}.

 \begin{figure}
  \centering
  \includegraphics[width=8cm]{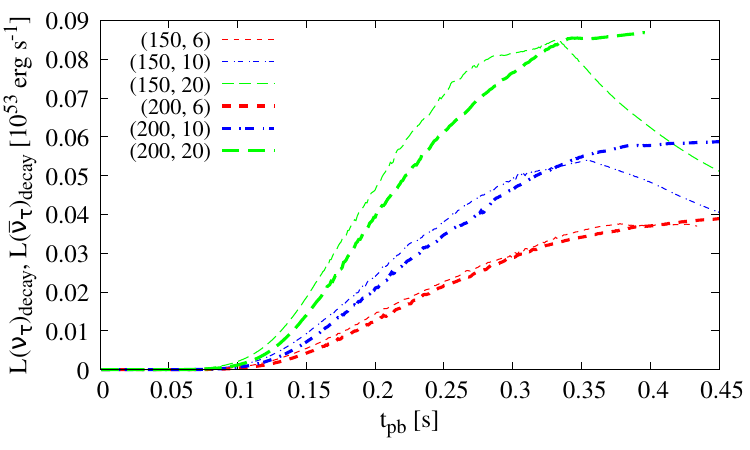}
  \caption{The luminosity of decay tau (anti)neutrinos as a function of $t_\mathrm{pb}$.  The thin lines are for the models with $m_\mathrm{s}=150$\,MeV and the thick lines are for the models with $m_\mathrm{s}=200$\,MeV.}
  \label{Lt}
 \end{figure}
  \begin{figure}
  \centering
  \includegraphics[width=8cm]{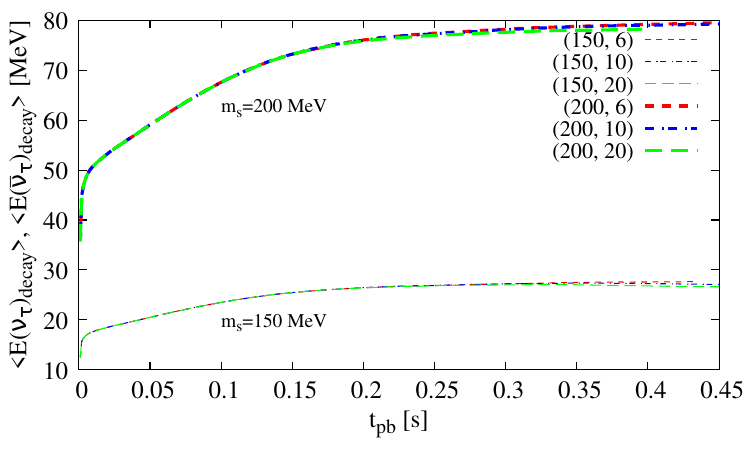}
  \caption{The mean energy of decay tau (anti)neutrinos as a function of $t_\mathrm{pb}$.  The thin lines are for the models with $m_\mathrm{s}=150$\,MeV and the thick lines are for the models with $m_\mathrm{s}=200$\,MeV, as shown in the labels.}
  \label{Et}
 \end{figure}

Since we consider sterile neutrinos that decay into tau neutrinos, there is a contribution of the daughter neutrinos to the neutrino luminosity. Figure~\ref{Lt} shows the luminosities of the tau neutrinos produced by sterile neutrino decay, $L_\mathrm{decay}$. In our parameter range, $L_\mathrm{decay}$ does not exceed the luminosity of tau neutrinos produced in the neutrinosphere, $L_\mathrm{SM}$. Contrary to $L_\mathrm{SM}$, $L_\mathrm{decay}$ increases as a function of the mixing angle because larger mixing angles lead to higher sterile neutrino fluxes. Compared with the large difference in the sterile neutrino luminosity  found in Fig.~\ref{Ls}, the difference of $L_\mathrm{decay}$ between the models with different sterile neutrino masses is less significant. This is because the fraction of the sterile neutrino energy which is distributed to tau neutrinos is different. The fraction is given by $1-x_{\gamma\gamma}\approx9.5$\% for $m_\mathrm{s}=150$\,MeV and $\approx27$\% for $m_\mathrm{s}=200$\,MeV. Thus, heavier sterile neutrinos give a larger fraction of their energy to tau neutrinos.

Since sterile neutrinos are  heavier than the mean energies for active neutrinos, decay tau neutrinos can be more energetic than the main component of the supernova neutrino signal. Fig.~\ref{Et} shows the mean energy of decay neutrinos, which is denoted as $\langle E_\mathrm{decay}\rangle$. One can see that $\langle E_\mathrm{decay}\rangle$ is sensitive to the sterile neutrino mass.  The mean energy reaches $\sim80$~MeV when $m_\mathrm{s}=200$~MeV, whereas $\langle E_\mathrm{decay}\rangle\sim28$~MeV when $m_\mathrm{s}=150$~MeV. This behavior of the mean energy can be explained as follows. The average Lorentz factor for sterile neutrinos is estimated as \cite{2009PhLB..670..281F,2018PhRvD..98j3010R},
 \begin{eqnarray}
     \gamma=1.3\left(\frac{T}{35\,\mathrm{MeV}}\right)^{0.4}\left(\frac{m_\mathrm{s}}{200\,\mathrm{MeV}}\right)^{-1},
 \end{eqnarray}
where $T$ is the temperature at the region where sterile neutrinos are produced. Thus, the mean energy of decay neutrinos can be roughly estimated as,
\begin{eqnarray}
    \langle E_\mathrm{decay}\rangle\approx\gamma m_\mathrm{s}(1-x_{\gamma\gamma})\approx130\,\mathrm{MeV}\left(1-\frac{m_\pi^2}{m_\mathrm{s}^2}\right),
\end{eqnarray}
where the weak temperature dependence is neglected. This expression shows that the mass dependence of $\langle E_\mathrm{decay}\rangle$ originates from the kinematics of the sterile neutrino decay. We consider the detactability of this higher energy neutrino flux next. 

\subsection{Event Number of Decay Neutrinos}

Although the luminosity $L_\mathrm{decay}$ of decay neutrinos is lower than $L_\mathrm{SM}$, their high mean energy for $m_\mathrm{s}=200$\,MeV could lead to distinguishable signatures in neutrino signals from a nearby supernova event \cite{2009PhLB..670..281F,2020JCAP...01..010M}. In this section, we estimate the electron anti-neutrino event number that could be detected by Hyper-Kamiokande \cite{2018arXiv180504163H,2021ApJ...916...15A}.

The neutrino flux observed on Earth would be affected by neutrino oscillation during propagation in stellar matter and vacuum. However, neutrinos in a dense environment could experience complicated processes such as collective oscillation \cite[e.g.,][]{2005NJPh....7...51B,2006PhRvL..97x1101D,Duan2010,Mirizzi2016} and the collisional flavor instability \cite[e.g.,][]{2023PhRvL.130s1001J,2023arXiv231005050L}, whose accurate treatment has not been established. We hence consider four phenomenological cases as shown in Tables \ref{tableOsc1} and \ref{tableOsc2}. 

In the ``No Oscillation" case, it is assumed that neutrinos do not oscillate during their propagation. In the ``NH" and ``IH" cases, we consider the matter (Mikheyev-Smirnov-Wolfenstein, MSW) effect \cite{1979PhRvD..20.2634W,1986NCimC...9...17M,PhysRevLett.56.1305}, assuming the normal and the inverted  hierarchy for the active neutrino masses, respectively. In the ``Swap" case, we assume that the  electron anti-neutrino flux on Earth is equal to the tau anti-neutrino flux on the neutrinosphere upto the geometrical factor.

The no oscillation case implies that the electron anti-neutrino flux on Earth is the same as the flux from the neutrinosphere up to the geometrical factor $1/D^2$, where $D$ is the distance to the event. In the ``NH" and ``IH" cases, active neutrinos undergo  flavor mixing induced by interactions with the surrounding matter during propagation. If we assume adiabaticity, this matter effect changes the neutrino flavors at the resonance density \cite[e.g.,][]{2000PhRvD..62c3007D,2006RPPh...69..971K},
\begin{eqnarray}
    \rho_\mathrm{res}&=&\frac{1}{2\sqrt{2}G_\mathrm{F}}\frac{\Delta m^2}{E}\frac{m_\mathrm{N}}{Y_e}\cos 2\theta\nonumber\\
    &\approx&1.3\times 10^2 \,\mathrm{g\,cm^{-3}}\times\nonumber\\
    &&\cos 2\theta\left(\frac{0.5}{Y_e}\right)\left(\frac{1\,\mathrm{MeV}}{E}\right)\left(\frac{\Delta m^2}{10^{-5}\,\mathrm{eV^2}}\right),
\end{eqnarray}
where $\Delta m^2$ is a mass squared difference, $E$ is the neutrino energy, $m_\mathrm{N}$ is the nucleon mass, $Y_e$ is the electron mole fraction, and $\theta$ is a mixing angle. It follows that, for the H-resonance, $\rho_\mathrm{res}\approx 3.9\times10^2$\,g\,cm$^{-3}$ for $m_\mathrm{s}=200$\,MeV and $\rho_\mathrm{res}\approx 1.1\times10^3$\,g\,cm$^{-3}$ for $m_\mathrm{s}=150$\,MeV, assuming $E\approx\langle E_\mathrm{decay}\rangle$. These densities correspond to the radius $r_\mathrm{res}\approx7.4\times10^4$\,km for $m_\mathrm{s}=200$\,MeV and $r_\mathrm{res}\approx5.3\times10^4$\,km for $m_\mathrm{s}=150$\,MeV. It then turns out that $r_\mathrm{res}$ is longer than the mean free path $\lambda_\mathrm{s}$ of sterile neutrinos with $m_\mathrm{s}=200$\,MeV, which is shown in Table I, whereas $r_\mathrm{res}$ can be shorter than $\lambda_\mathrm{s}$ when $m_\mathrm{s}=150$\,MeV. Thus, in the case of the inverted mass hierarchy, almost all decay tau antineutrinos would undergo the H-resonance when $m_\mathrm{s}=200$\,MeV, whereas a part of the decay antineutrinos can evade the H-resonance when $m_\mathrm{s}=150$\,MeV. On the other hand, we do not consider the effect of the L-resonance because it does not affect the antineutrino sector \cite{2000PhRvD..62c3007D,2006RPPh...69..971K}.

Considering the above estimates, we calculate the electron anti-neutrino flux on the Earth as follows in the ``NH" and ``IH" cases. In the case of the normal mass hierarchy, tau antineutrinos do not experience the MSW resonance. The electron anti-neutrino flux on Earth can then be estimated as \cite{2015PhRvD..91f5016W,2020PhRvD.101f3027S},
\begin{eqnarray}
f(\bar{\nu}_e)&=&c_{12}^2c_{13}^2f_\mathrm{SM}(\bar{\nu}_e)+s_{12}^2c_{13}^2(f_\mathrm{SM}(\bar{\nu}_x)+s_{23}^2f_\mathrm{decay}(\bar{\nu}_\tau))\nonumber\\
&&+s^2_{13}(f_\mathrm{SM}(\bar{\nu}_y)+c_{23}^2f_\mathrm{decay}(\bar{\nu}_\tau))\nonumber\\
&\approx& c_{12}^2f_\mathrm{SM}(\bar{\nu}_e)+s_{12}^2(f_\mathrm{SM}(\bar{\nu}_x)+s_{23}^2f_\mathrm{decay}(\bar{\nu}_\tau))
\end{eqnarray}
where $\nu_x$ and $\nu_y$ are rotated basis defined as $(|\nu_e\rangle,\;|\nu_x\rangle,\;|\nu_y\rangle)=R^{-1}_{23}(\theta_{23})(|\nu_e\rangle,\;|\nu_\mu\rangle,\;|\nu_\tau\rangle)$, $R_{23}$ is the rotation matrix in the 2-3 space, $f_\mathrm{SM}$ is the flux of neutrinos produced on the neutrinosphere,  $f_\mathrm{decay}$ is the flux of decay neutrinos, and $c_{ij}=\cos\theta_{ij}$ and $s_{ij}=\sin\theta_{ij}$ for integers $i$ and $j$. In the case of the inverted hierarchy,
\begin{eqnarray}
f(\bar{\nu}_e)&=&
s_{13}^2(f_\mathrm{SM}(\bar{\nu}_e)+c_{23}^2c_{13}^2f_\mathrm{decay}^\mathrm{out}(\bar{\nu}_\tau))\nonumber\\
&&+s_{12}^2c_{13}^2(f_\mathrm{SM}(\bar{\nu}_x)+s_{23}^2f_\mathrm{decay}^\mathrm{in}(\bar{\nu}_\tau)\nonumber\\
&&+|-c_{12}s_{23}-s_{12}c_{23}s_{13}e^{i\delta}|^2f_\mathrm{decay}^\mathrm{out}(\bar{\nu}_\tau))\nonumber\\
&&+c^2_{12}c^2_{13}(f_\mathrm{SM}(\bar{\nu}_y)+c_{23}^2f_\mathrm{decay}^\mathrm{in}(\bar{\nu}_\tau)\nonumber\\
&&+|s_{12}s_{23}-c_{12}c_{23}s_{13}e^{i\delta}|^2f_\mathrm{decay}^\mathrm{out}(\bar{\nu}_\tau))\label{f_IH}\\
&\approx&s_{12}^2(f_\mathrm{SM}(\bar{\nu}_x)+s_{23}^2f_\mathrm{decay}^\mathrm{in}(\bar{\nu}_\tau)+c_{12}^2s_{23}^2f_\mathrm{decay}^\mathrm{out}(\bar{\nu}_\tau))\nonumber\\
&&+c^2_{12}(f_\mathrm{SM}(\bar{\nu}_y)+c_{23}^2f_\mathrm{decay}^\mathrm{in}(\bar{\nu}_\tau)+s_{12}^2s_{23}^2f_\mathrm{decay}^\mathrm{out}(\bar{\nu}_\tau)),\nonumber
\end{eqnarray}
where $\delta$ is the CP violation phase and $f^\mathrm{in}_\mathrm{decay}(\bar{\nu}_\tau)$ and $f^\mathrm{out}_\mathrm{decay}(\bar{\nu}_\tau)$ are the decay neutrino fluxes from the region where $r<r_\mathrm{res}$ and $r>r_\mathrm{res}$, respectively. In Eq.~(\ref{f_IH}), we assume the incoherent vacuum oscillation of decay neutrinos produced at $r>r_\mathrm{res}$. 

\begin{table}[]
\begin{tabular}{c|cccc}
Model&No Oscillation& NH&IH&Swap\\\hline\hline
   Standard  &   30539   & 30698 & 31018& 31030                           \\\hline

$(150,\;6)$      & 29666 &      30014 &        30714      &  30739        \\
$(150,\;10)$       & 29803&     29981&    30338&  30350 \\
$(150,\;20)$        & 28436&        28793&       29513&        29538     \\
\hline
$(200,\;6)$       & 29408&        29774&      30513&        30539      \\
$(200,\;10)$&29142&      29510&30251&      30278                                    \\
$(200,\;20)$       &27038   &  27851 &   29489  &  29547    \\

\end{tabular}
\caption{The number of neutrino events $N_\mathrm{SM}$ integrated from $t_\mathrm{pb}=0$\,s to 0.4\,s from a supernova at $D=8.5$\,kpc observed by Hyper-Kamiokande. Here, neutrinos only from the neutrinosphere are counted (see Table \ref{tableOsc2} for the contribution from sterile neutrino decays). We consider four cases for different treatments of the neutrino oscillation. In the ``No oscillation" case, oscillation is not implemented. In the ``NH" and ``IH" cases, the MSW effect is taken into account, assuming the normal and the inverted mass hierarchies, respectively. In the ``Swap" case, the complete flavor swap between $\bar{\nu}_e$ and $\bar{\nu}_\tau$ is assumed.}
\label{tableOsc1}
\end{table}

\begin{table}[]
\begin{tabular}{c|cccc}
Model&No Oscillation& NH&IH&Swap\\\hline\hline
   Standard  &   0&0&0&0                          \\\hline

$(150,\;6)$      & 0&   274        &   553                                 &   1588       \\
$(150,\;10)$       & 0 &     429                             &  1045&  2481  \\
$(150,\;20)$        & 0 &    705                                  &  1949&  4077       \\
\hline
$(200,\;6)$       & 0 &   794                               & 2196&4594           \\
$(200,\;10)$&0 & 1279        & 3533&    7391                                    \\
$(200,\;20)$       &0 &   2058                                & 5686 &  11897    \\

\end{tabular}
\caption{The same as Table II, but for the number of neutrino detections from sterile neutrino decay, $N_\mathrm{decay}$.}
\label{tableOsc2}
\end{table}

Tables II and III show the electron anti-neutrino event numbers integrated over $t_\mathrm{pb}=0$--0.4\,s observed by Hyper-Kamiokande through the inverse $\beta$-decay. Here we use $s_{12}^2=0.310$ and $s_{23}^2=0.558$ \cite{2019JHEP...01..106E} and ignore the CP violation phase. We assume a supernova event at the Galactic center, whose distance to the Solar System is $D=8.5$\,kpc. The volume of Hyper-Kamiokande is assumed to be 220\,kt and the detection threshold neutrino energy is assumed to be 8.3\,MeV.  The spectrum of neutrinos  is fitted by a function \cite{2003ApJ...590..971K},
\begin{eqnarray}
     f(E)=\frac{(1+\alpha)^{(1+\alpha)}}{\Gamma(1+\alpha)}\frac{E^\alpha}{\langle E\rangle^{\alpha+1}}\exp\left(-(1+\alpha)\frac{E}{\langle E\rangle}\right),
\end{eqnarray}
where $\alpha=(\langle E^2\rangle-2\langle E\rangle)/(\langle E\rangle-\langle E^2\rangle)$, to estimate the number $N_\mathrm{SM}$ of neutrinos from the neutrinosphere. For the decay neutrino number $N_\mathrm{decay}$, we treat decay neutrinos as monoenergetic particles with mean energy shown in Fig.~\ref{Et}.

The tables show that when the mixing angle is larger, $N_\mathrm{SM}$ becomes smaller because the active neutrino luminosity is decreased by sterile neutrinos. On the other hand, the contribution of decay neutrinos increases as the mixing angle becomes larger because of the higher sterile neutrino luminosity. Furthermore, the contribution of  $N_\mathrm{SM}$ to the total neutrino event number is larger than $N_\mathrm{decay}$ in the parameter region discussed in this study. In the ``No Oscillation" case, decay neutrinos cannot be detected using the inverse $\beta$ decay because they are tau antineutrinos. On the other hand, the expected event number becomes the largest in the ``Swap" case because all of the decay neutrinos are swapped with electron antineutrinos. Although the accurate treatment for the neutrino oscillation is not established, the ``Swap" case works as an upper limit on the decay neutrino event number. 

\begin{figure}
  \centering
  \includegraphics[width=8cm]{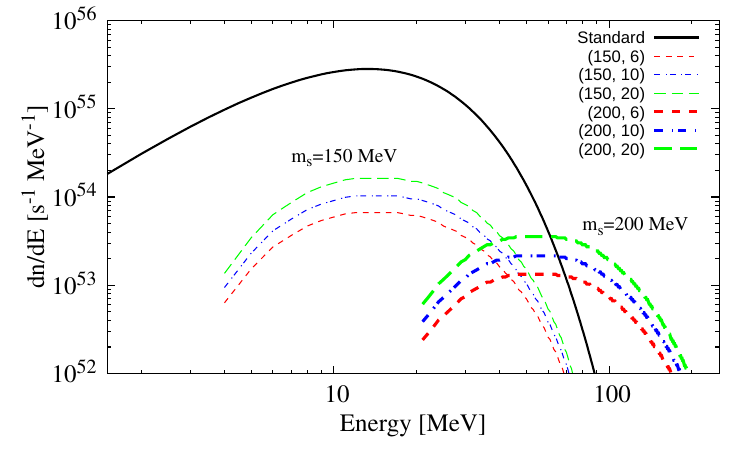}
  \caption{The number spectra $dn/dE$ of tau antineutrinos at $t_\mathrm{pb}=0.35$\,s. The solid line shows the component produced on the neutrinosphere in the ``Standard" model, and the broken lines show the contribution of decay antineutrinos.  The thin lines are for the models with $m_\mathrm{s}=150$\,MeV and the thick lines are for the models with $m_\mathrm{s}=200$\,MeV, as shown in the labels.}
  \label{spec_decay}
 \end{figure}
 \begin{figure}
  \centering
  \includegraphics[width=8cm]{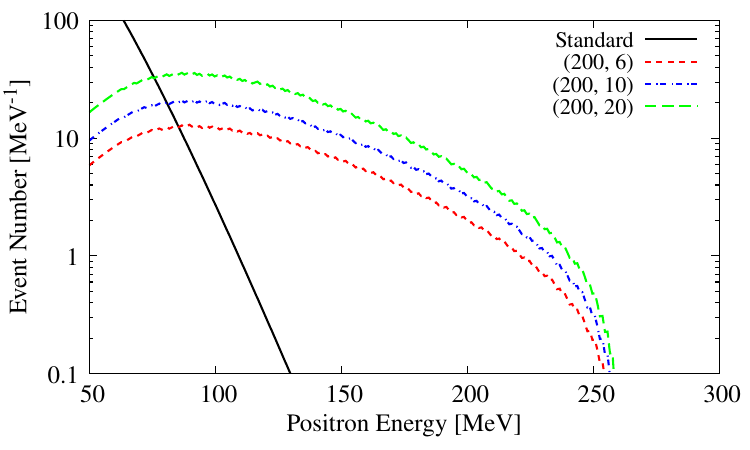}
  \caption{The neutrino event number observed by Hyper-Kamiokande as functions of the positron energy in 1\,MeV bins, integrated over $t_\mathrm{pb}=0$--0.4\,s. The supernova is assumed to be located at $D=8.5$\,kpc and the ``Swap" case is adopted to take into account neutrino oscillations.  The solid line shows the component produced on the neutrinosphere in the ``Standard" model, and the broken lines show the contribution of decay antineutrinos in the models with $m_\mathrm{s}=200$\,MeV.}
  \label{event_num}
 \end{figure}

A key distinguishing feature of the decay neutrinos is its higher energy. As shown in Fig.~\ref{Et}, when $m_\mathrm{s}=200$\,MeV, $\langle E_\mathrm{decay}\rangle$ reaches 80\,MeV, which is much higher than the mean energy of neutrinos from the neutrinosphere. This leads to a possibility that decay neutrinos could form a distinguishable bump in the neutrino spectrum \cite{2020JCAP...01..010M}. Fig.~\ref{spec_decay} shows the number spectra of tau antineutrinos that are produced by the sterile neutrino decay (shown in the broken lines) and standard processes (shown in the solid line) at $t_\mathrm{pb}=350$\,ms. Here, instead of the fitting formula Eq.~(\ref{Qcool}) used in the simulations to reduce computational cost, we evaluate the collisional integral in order to explore the spectra of decay neutrinos, 
\begin{eqnarray}
    \mathcal{C}_\mathrm{coll}=\frac{1}{2E_\mathrm{s}}\int d^3\hat{p}_2d^3\hat{p}_3d^3\hat{p}_4(1-f_2)f_3f_4\nonumber\\
    \times S|M|^2\delta^{(4)}(p_\mathrm{s}+p_2-p_3-p_4)(2\pi)^4,
\end{eqnarray}
where $E_\mathrm{s}$ is the sterile neutrino energy, $p_\mathrm{s}$, $p_2$, $p_3$, and $p_4$ are the momenta of neutrinos involved in the reactions, $f_2$, $f_3$, and $f_4$ are the Fermi-Dirac distributions of active neutrinos, $S$ is the symmetry factor, and $|M|^2$ is the amplitude tablated in Refs.~\cite{2009PhLB..670..281F,2020JCAP...01..010M}. We adopt a reduction method developed in Refs.~\cite{1976Ap&SS..39..429Y,1995PhRvD..52.1764H,2021PhRvD.104a6026M} to calculate the integral. We then estimate the decay neutrino spectrum as \cite{2020JCAP...01..010M},
\begin{eqnarray}
    \frac{dn}{dE}=\frac{m_\mathrm{s}}{2\bar{E}}\int^\infty_{E_\mathrm{min}}dE_\mathrm{s}\frac{1}{p_\mathrm{s}}\frac{dn_\mathrm{s}}{dE_\mathrm{s}},
\end{eqnarray}
where $p_\mathrm{s}$ is the sterile neutrino momentum, $dn_\mathrm{s}/dE_\mathrm{s}$ is the sterile neutrino spectrum obtained from the collisional integral, $\bar{E}=(m_\mathrm{s}^2-m_\pi^2)/2m_\mathrm{s}$, and $E_\mathrm{min}=m_\mathrm{s}(E^2+\bar{E}^2)/2E\bar{E}$.

Figure~\ref{spec_decay} shows that decay antineutrinos form a high-energy bump at $E\sim60$\;MeV in the cases of $m_\mathrm{s}=200$\,MeV, while the bump is located at $E\sim20$\,MeV in the $m_\mathrm{s}=150$\,MeV cases. Although the total event number of the decay component observed by water-Cherenkov detectors will be smaller than the standard component as shown in Tables II and III, the bump could appear in the spectrum as a detectable high-energy component at $E\gtrsim60$\,MeV if $m_\mathrm{s}\sim200$\,MeV. On the other hand, when sterile neutrinos are lighter, it would be difficult to distinguish the signals of decay neutrinos from the standard component.

Figure \ref{event_num} shows the expected neutrino event number per 1\,MeV observed by Hyper-Kamiokande as a function of the positron energy. The supernova event is assumed to be located at $D=8.5$\,kpc and the neutrino oscillation is treated with the ``Swap" scheme, which gives an upper limit on the electron antineutrino flux on Earth. One can see that, if we adopt the energy bin width of $\sim10$\,MeV, the decay antineutrinos up to $E\sim250$\,MeV would be above the detection threshold. This implies that the high-energy tail of the neutrino spectrum will play a critical role in investigating a signature of sterile neutrinos. We note, however, that high-energy neutrinos could be produced non-thermally by shock acceleration even without sterile neutrinos \cite[e.g.,][]{1989ApJ...342..416G,2021MNRAS.502...89N}. Nevertheless, given the typically power-law nature of non-thermal emission, the spectra could be a distinguishing feature. Since detailed studies on the shock acceleration with multidimensional supernova models have been yet to be conducted, we leave a detailed comparison between the non-thermal component and the non-standard decay component for future study. 

\subsection{Gravitational Waves}

\begin{figure}
  \centering
  \includegraphics[width=8cm]{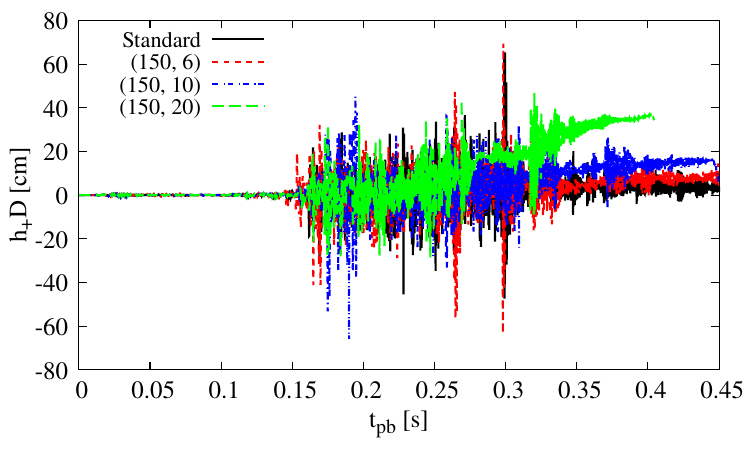}
   \includegraphics[width=8cm]{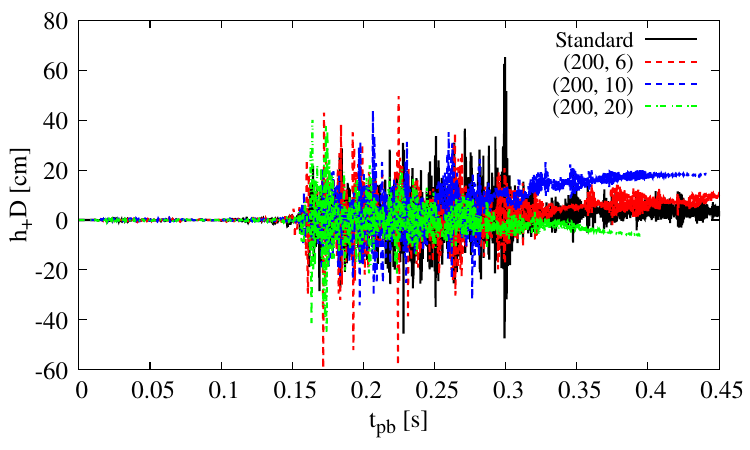}
  \caption{The strain $h_+$ of the GW plus mode times the distance $D$ to the supernova event as a function of  $t_\mathrm{pb}$. The upper panel is for the models with $m_\mathrm{s}=150$\,MeV and the lower panel is for the ones with $m_\mathrm{s}=200$\,MeV. The solid line shows the result for the ``Standard" model  and the others are for the models with sterile neutrinos.}
  \label{GW}
 \end{figure}

Multi-dimensional supernova models have indicated that a nearby supernova event can be a source of gravitational waves (GWs; \cite[e.g.,][]{2006RPPh...69..971K}), although they have not been detected yet. GWs can provide information on the supernova core because of their feeble interaction with matter.

 \begin{figure}
  \centering
  \includegraphics[width=8cm]{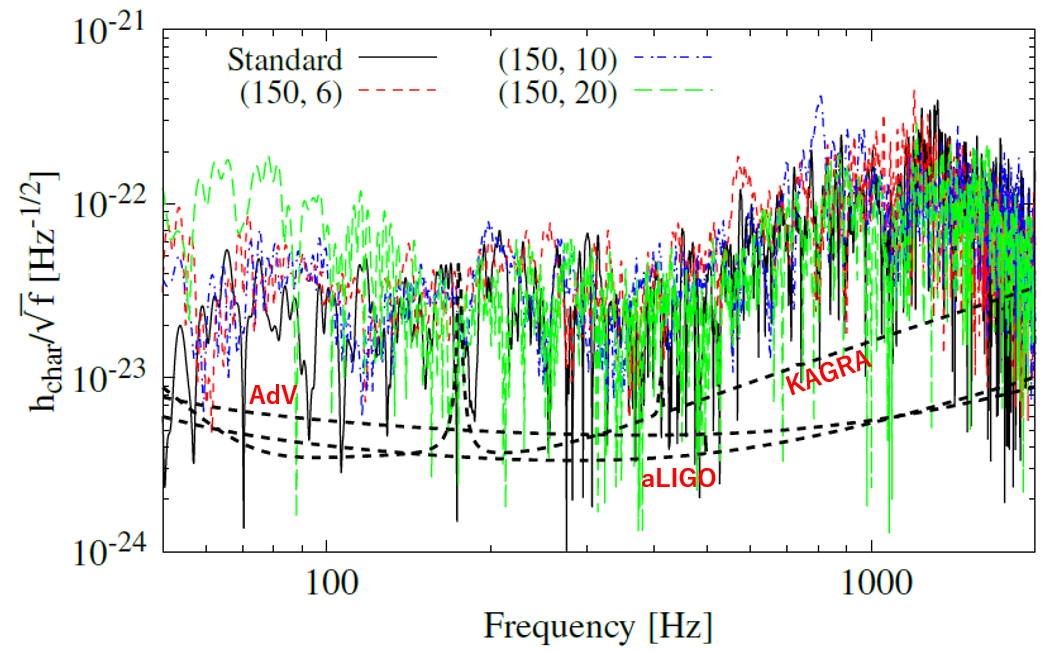}
   \includegraphics[width=8cm]{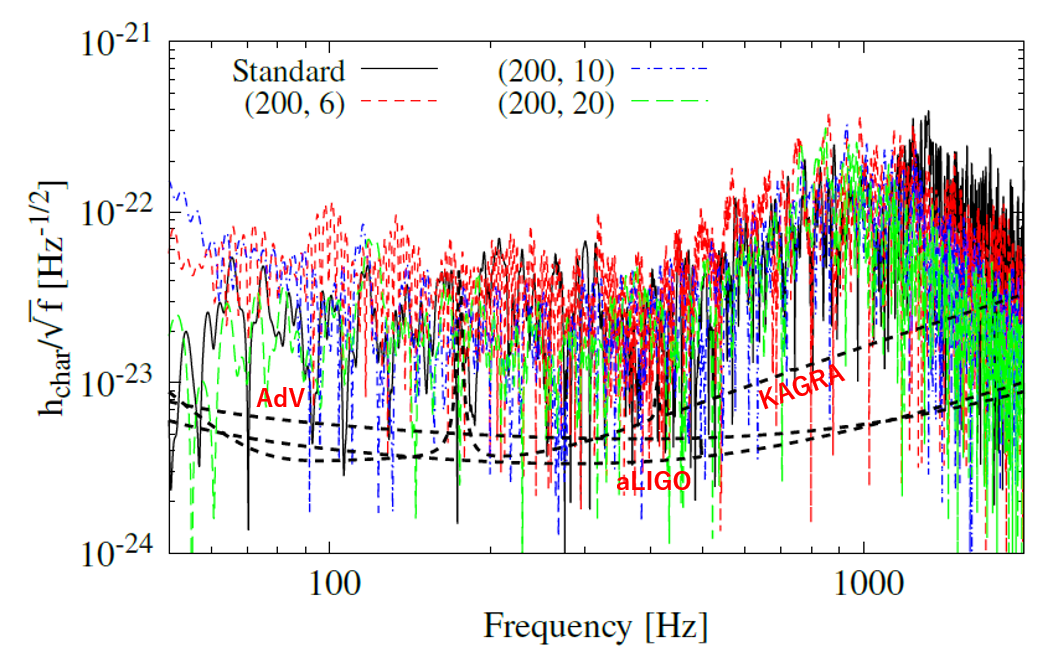}
  \caption{The characteristic GW strain defined in Eq.~(\ref{spec}) divided by $\sqrt{f}$ from a supernova event at $D=8.5$\,kpc. Sensitivities of Advenced LIGO \cite{aLIGO}, Advanced VIRGO \cite{AdV}, and KAGRA \cite{KAGRA} are also shown. The upper panel is for the models with $m_\mathrm{s}=150$\,MeV and the lower panel is for the ones with $m_\mathrm{s}=200$\,MeV. The solid line shows the result for the ``Standard" model  and the others are for the models with sterile neutrinos.}
  \label{GW_spec}
 \end{figure} 

In general, GWs can be decomposed into two modes, the plus-mode and the cross-mode. Axisymmetric models can only predict the plus-mode. Fig.~\ref{GW} shows the strain of the GW plus-mode observed from the equatorial direction. In all of the models, the amplitude of the GW strain $h_+$  is tiny until $t_\mathrm{pb}\sim0.15$\,s because the mass accretion is spherically symmetric. However, it becomes larger after the phase because asymmetric motion behind the bounce shock starts growing. We can see that the strain becomes smaller when the mixing angle is larger. This is because the mass accretion on a proto-neutron star is suppressed by the additional heating. Also, when the mixing angle is sufficiently large, the GW waveform deviates from 0. Except for the $(200,\;20)$ model, the strain maintains positive values. This memory effect can be attributed to the asymmetric expansion of matter \cite{2009ApJ...707.1173M}. The positive deviation implies a prolate explosion, whereas the negative deviation implies a oblate explosion. Since the matter tends to expand towards the polar direction, the explosion often becomes prolate.

 \begin{figure}
  \centering
  \includegraphics[width=8cm]{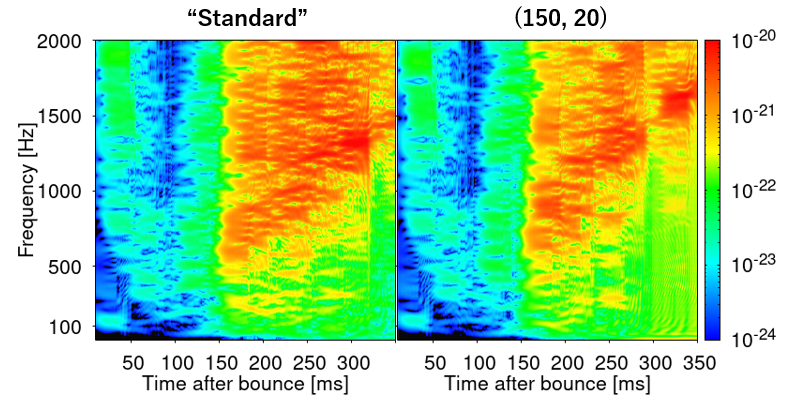}
  \caption{The spectrograms for the ``Standard" and $(150,\;20)$ models, where the color shows the value of $h_\mathrm{char}$. The supernova event is assumed to be located at $D=8.5$\,kpc.}
  \label{spectrogram}
 \end{figure} 

Figure \ref{GW_spec} shows the characteristic strain defined as,
\begin{eqnarray}
    h_\mathrm{char}=\sqrt{\frac{2}{\pi D^2}\frac{dE_\mathrm{GW}}{df}},\label{spec}
\end{eqnarray}
where $dE_\mathrm{GW}/df$ is the GW spectral energy density. In this figure, $D$ is fixed to 8.5\,kpc. All models show a peak at $f\sim1$\,kHz. In a lower frequency region at $f\sim100$\,kHz, the $(150,\;20)$ model shows an enhancement compared with the other models. This is because the model produces the highest luminosity of sterile neutrinos, as shown in Fig.~\ref{Ls}, and hence causes the most significant memory effect. Fig.~\ref{GW_spec} shows the sensitivity of Advanced LIGO \cite{aLIGO}, Advanced VIRGO \cite{AdV}, and KAGRA \cite{KAGRA} as well. It can be seen that the GW signals would be observable in our parameter range if a supernova event appears at the Galactic center.

Figure \ref{spectrogram} shows the spectrograms for the ``Standard" and $(150,\;20)$ models. One can see that the  two spectrograms are similar to each other and it would be difficult to distinguish them observationally. In general, the GW spectrum is dependent on the PNS mass and radius. However, the effects of sterile neutrinos are too small to cause qualitative differences in the GW spectrum in our parameter region.


\section{Discussion}\label{ref:discuss}

In this study, we performed two-dimensional simulations of stellar core collapse including the transport of heavy sterile neutrinos. We found that photons produced through the decay channel $\nu_s\rightarrow\nu_\tau\pi^0\rightarrow\nu_\tau\gamma\gamma$ can heat the stellar material to enhance the explosion energy and the synthesized nickel mass. In addition, sterile neutrinos can suppress active neutrino and GW signals from a nearby supernova event, although they could be still observed if the event appears at the Galactic center. Interestingly, the additional cooling induced by the sterile neutrino production led to the proto-neutron star contraction and increased the active neutrino mean energy when $m_\mathrm{s}=150$\,MeV. It was also found that decay tau neutrinos have higher mean energies than neutrinos from the neutrinosphere when $m_\mathrm{s}=200$\,MeV. In this case, it would be possible to detect a high energy bump in neutrino spectra by, e.g., Hyper-Kamiokande, which would be a signature of sterile neutrinos.

It has been pointed out that one of the most useful observables to constrain feebly-interacting particles that are produced in supernovae is the explosion energy \cite{2019PhRvD..99l1305S,2022PhRvD.105f3009M,2022PhRvL.128v1103C}. Since the explosion energy for our sterile neutrino models reach $\sim10^{51}$\,erg and are still increasing at the end of the simulations, the sterile neutrino parameters with $\sin^2\theta_{\tau4}\gtrsim1\times10^{-7}$ for $m_\mathrm{s}\sim150$--200\,MeV would presumably be excluded. Recently, this argument was applied to the explosion energies of low-energy supernova events to obtain tight constraints \cite{2023arXiv230905860C,2023arXiv231100033C}. Although the parameter range in this study is in the excluded region based on the explosion energy, our prediction on multi-messenger signals from a nearby supernova event would work as a method to obtain an independent constraint. Apart from the explosion energy, we showed the effects of sterile neutrinos on the thermal component of active neutrinos and GWs in Section III. Although the signals are slightly suppressed, their variations are within model uncertainties. We hence conclude that the high-energy bump in the neutrino spectrum and the enhanced explosion energy would be particularly useful for the search of heavy sterile neutrinos.

It is known that two- and three-dimensional supernova models can behave in qualitatively different ways \cite[e.g.,][]{2014ApJ...785..123C,2014ApJ...786...83T,2017MNRAS.472..491M,2018ApJ...855L...3O,2020MNRAS.491.2715B,2020ApJ...896..102K,2021ApJ...915...28B}, because turbulence cascades to small scales in three-dimensional models, whereas it cascades to large scales in two-dimensional models. In addition, two-dimensional models cannot predict the cross-mode of GWs. More sophisticated treatment of general relativity (GR) would be indispensable to obtain temperature of proto-neutron stars \cite{Muller2010,Rahman2022,Kuroda2023,2020ApJ...896..102K}.
It is hence desirable to perform three-dimensional GR simulations that takes sterile neutrinos into account to provide more accurate predictions. It would also be important to improve the transport scheme for sterile neutrinos. In our simulations, the transport is treated with the ray-by-ray approximation, in which sterile neutrinos propagate only in the radial direction. Although it could be computationally expensive, improving the transport scheme would lead to more accurate results.

Finally, in this study we have not taken into account detailed neutrino oscillation, e.g., fast flavor conversion
\cite{Tamborra2021,Capozzi2022review,Richers2022review,Volpe2023review} and collisional flavor instability \cite{2023PhRvL.130s1001J}. Those flavor conversion may affect the neutrino heating rate and explodability of supernovae \cite{Suwa_2011,Ehring2023a,Ehring2023b,Nagakura2023may}. In strongly magnetized proto-neutron stars, spin-flavor precession would mix the flavors \cite{Sasaki2021,Sasaki2023}. Neutrinos also may lead to chiral plasma instability \cite{Masada2018,Matsumoto2022,Kamada2023}. These ingredients should be considered in the future to update the ``Standard'' model.

In summary, in this work we focused on heavy sterile neutrinos with $m_\mathrm{s}\sim200$\,MeV in the context of core-collapse supernovae. However, the parameter space for the sterile neutrino particle is vast, and  different mass can cause various phenomena in supernovae \cite{1991NuPhB.358..435K,1992A&A...254..121P,1993APh.....1..165R,1997PhRvD..56.1704N,1999PhRvC..59.2873M,2000PhRvD..61l3005C,2003APh....18..433F,2006PhRvD..73i3007B,2007PhRvD..76l5026K,2012JCAP...01..013T,2014PhRvD..89f1303W,1994PhLB..323..360S,2001PhRvD..64b3501A,2006PhRvD..74l5015H,2007PhRvD..76h3516H,2011PhRvD..83i3014R,2014PhRvD..90j3007W,2016IJMPA..3150137W,2019PhRvD..99d3012A,2019JCAP...12..019S,2020JCAP...08..018S,2022PhRvD.106a5017S}. Such phenomenology has been investigated only with one-dimensional supernova models. Although one can estimate sterile neutrino luminosities from a supernova core with one-dimensional models, signatures of sterile neutrinos in multi-messenger signals from supernova events cannot be predicted without multi-dimensional simulations. It is hence desirable to extend this study to a wider range of the parameter space to fully utilize supernovae as a laboratory for sterile neutrinos.

\begin{acknowledgments}
We thank Garv Chauhan for discussions. Numerical computations were  carried out on Cray XC50 at Center for Computational Astrophysics, National Astronomical Observatory of Japan. This work is supported by Research Institute of Stellar Explosive Phenomena at Fukuoka University and JSPS KAKENHI Grant Numbers JP21H01088, JP22H01223, JP23H01199, JP23KJ2147, JP23K03400 and JP23K13107. This work is supported in part by funding from Fukuoka University (Grant No.GR2302).
This research was also supported by MEXT as “Program for Promoting researches on the Supercomputer Fugaku” (Structure and Evolution of the Universe Unraveled by Fusion of Simulation and AI; Grant Number JPMXP1020230406) and JICFuS.
The work of SH is supported by the U.S.~Department of Energy Office of Science under award number DE-SC0020262, NSF Grant No.~AST1908960 and No.~PHY-2209420, and JSPS KAKENHI Grant Number JP22K03630 and JP23H04899. This work was supported by World Premier International Research Center Initiative (WPI Initiative), MEXT, Japan.
\end{acknowledgments}

\bibliography{ref.bib}
\end{document}